\documentclass[preprint,showpacs,groupedaddress,amsmath,amssymb]{revtex4-1}  
\usepackage{graphicx,color}
\usepackage{multirow}
\begin{document}
\title{ Electronic structure of interfaces between hexagonal 
and rhombohedral graphite }
\author{ M. Taut, and K. Koepernik }
\affiliation{\small IFW Dresden, PO Box 270116, D-01171 Dresden, Germany}
\date{{\small (\today)}}

\begin{abstract}
  An analysis of the electronic structure
    of interfaces between hexagonal (AB) and rhombohedral
    (ABC) graphite based on density functional theory is presented.
    Both of the two simplest interface structures host
    (localized) interface bands, which are located around the K-point
    in the Brillouin zone, and which give rise to strong peaks in the
    density of states at the Fermi level.  All interface bands near the Fermi
    energy are localized at monomers (single atoms with dangling $p_z$
    orbitals), whereas those around 0.5 eV belong to $p_z$-bonded
    trimers, which are introduced by the interface and which are not
    found in the two adjacent bulk substances.  There is also an
    interface band at the (AB) side of the interface which resembles
    one of the interface states near a stacking fault in (AB)
    graphite.
\end{abstract}


\maketitle


\section{ Introduction}
Stacking faults (SFs) in graphitic  stacks play an important role for the
electronic properties, because they induce localized bands
and sharp peaks in the local density of states (LDOS) near
the SFs and the Fermi energy $\varepsilon_F$.
 This is especially important because
the two competing periodic {\em bulk} structures 
(hexagonal (AB) and rhombohedral (ABC)
 graphite) have a very small intrinsic density of states (DOS) at
 $\varepsilon_F$ \cite{taut13,taut14}.
Therefore, the stacking faults produce
 virtually two-dimensional (2D) electron gases
with little coupling to the bulk states and
 a thickness (extent) of a few lattice constants.

In a series of 3 papers we investigated the electronic structure of
stacking faults, restricting ourselves to the simplest possibility
where the SF separates two monolithic bulk blocks, as well as the cases
of single displaced surface layers.  In principle, there are 3
highly symmetric possibilities:
both blocks can be hexagonal \cite{taut13}, or rhombohedral \cite{taut14},
or one block is hexagonal and the other one is rhombohedral (this paper).
In each case there are two possibilities (denoted by $\alpha$ and $\beta)$,
 depending on whether the shift between
the two blocks is +1/3 or -1/3 of the lateral lattice constant.
For displaced surface layers on hexagonal \cite{taut13}
or rhombohedral \cite{taut14} graphite there is
only one possibility in either case.
The intriguing new feature in interfaces between hexagonal and
rhombohedral graphite is that  (semi)localized bands are possible 
which decay into one, or the other, or into both bulk materials.

For the one-particle spectra the
generalized gradient approximation (GGA) \cite{PBE96}
has been used, whereas 
total energies are calculated with the local density approximation 
(LDA). 
The reason for this choice has been discussed in detail in  \cite{taut13}.
are compared with those from the LDA.
 The calculation of the electronic properties was done with the
Full-Potential Local-Orbital  DFT package (FPLO) 
\cite{fplo,koepernik99a, koepernik99b} using default settings except
for the $k$-mesh of the Brillouin zone (BZ) integration, which is
specified at the appropriate places in the text below.

\begin{figure}[h]
   \centering
   \includegraphics[width=0.15\textwidth]{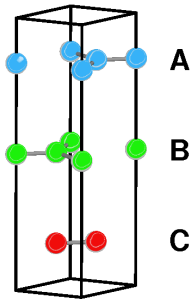}
   \caption{ (Color online) The three basic layers in graphitic stacks within 
a hexagonal unit cell: A(blue), B(green), and C(red).
} 
   \label{fig:atoms-ABC}
\end{figure}


\section{Projected bulk band structure }

\begin{figure}[h]
   \centering
 \includegraphics[width=0.45\textwidth]{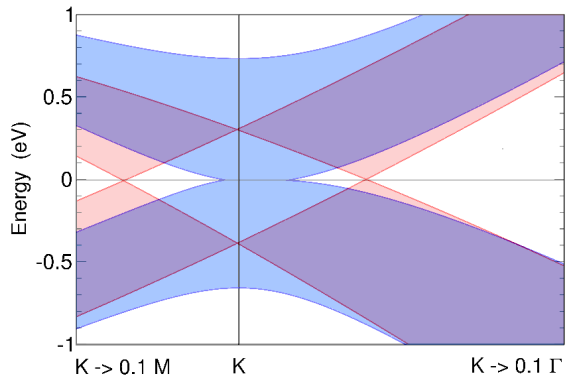}
 \caption{ (Color online) Band structure of bulk (AB) and (ABC)-graphite 
  projected onto a plane parallel to the layers shown in 
blue (covering the K-point) and red (avoiding the K point) shades, 
respectively. 
Only 10 \% of the symmetry lines $M-K$ and $K-\Gamma$
centered around symmetry point K, are shown. 
The end points of this region are labeled with $K \to 0.1 M$ and $K \to 0.1 \Gamma$. 
}
   \label{fig:projBS}
\end{figure}

The projected bulk band structure (PBBS) is a very helpful device
 for the interpretation of calculations of the electronic band structure of 
interfaces. It consists of broad quasi-continuous bands which comprise 
the electronic bands with periodic boundary conditions as 
a function of $k_x, k_y$  for quasi-continuous $k_z$ as parameters,
 where $x,y$ are the 
Cartesian coordinates parallel to the interface 
and $z$ is perpendicular to it. 
The bulk band structures of $(AB)$ and $(ABC)$ graphite can be found in 
Refs.\cite{taut13}  and \cite{taut14}, respectively, and the PBBSs 
 around the K-point are shown in Fig.\ref{fig:projBS}
 as shades of different color in the same plot. 
Evanescent states for 
each $k_x$ and $k_y$ can exist in each material 
only in {\em gaps} of the corresponding PBBS \cite{heine63,heine64},
 because in the {\em bulk state continuum}
an evanescent state could mix with a bulk state (of the same energy and $k_x$ and $k_y$) 
 and thus  turn into a bulk state or interface resonance. 

 Fig.\ref{fig:projBS} shows the regions where states can exist which 
fall off  into the (AB) bulk, into the (ABC) bulk,  and into both 
 sides. Obviously, at the very K point only boundary states which 
fall off into the (ABC) bulk are allowed, but there are windows 
on both symmetry lines (K-M and K-$\Gamma$) near the K point where 
real localized states can occur. 
We have to consider, however, that this rule is exactly valid only 
for interfaces of two infinite (or sufficiently thick) 
blocks of different materials.
In case of any doubt one has to analyze the character of the wave functions, 
e.g. by calculating the band weights (see Ref.\cite{taut13}).

\section{ Types of Interfaces}

\begin{figure}[h]
\begin{minipage}{.45\textwidth}
\includegraphics[width=0.7\textwidth]{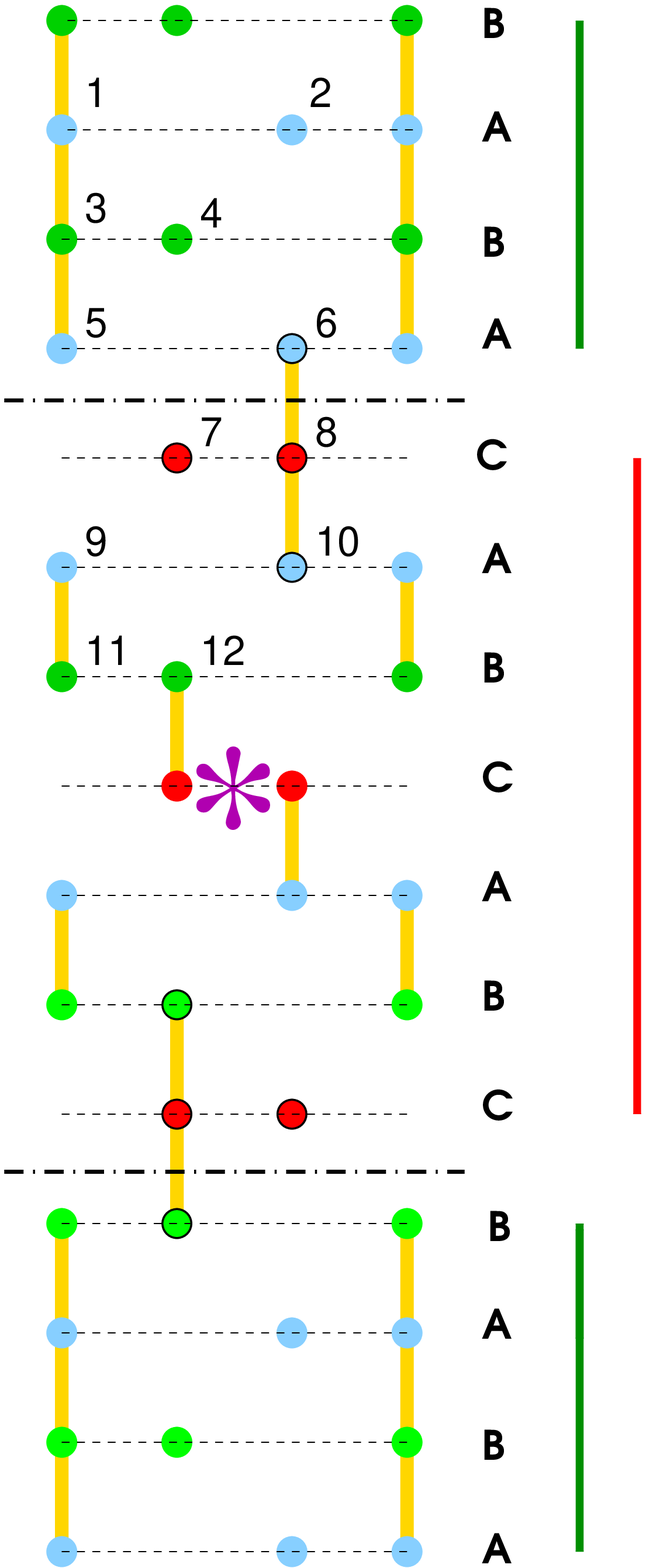}
\end{minipage}
\begin{minipage}{0.45\textwidth}
\includegraphics[width=0.7\textwidth]{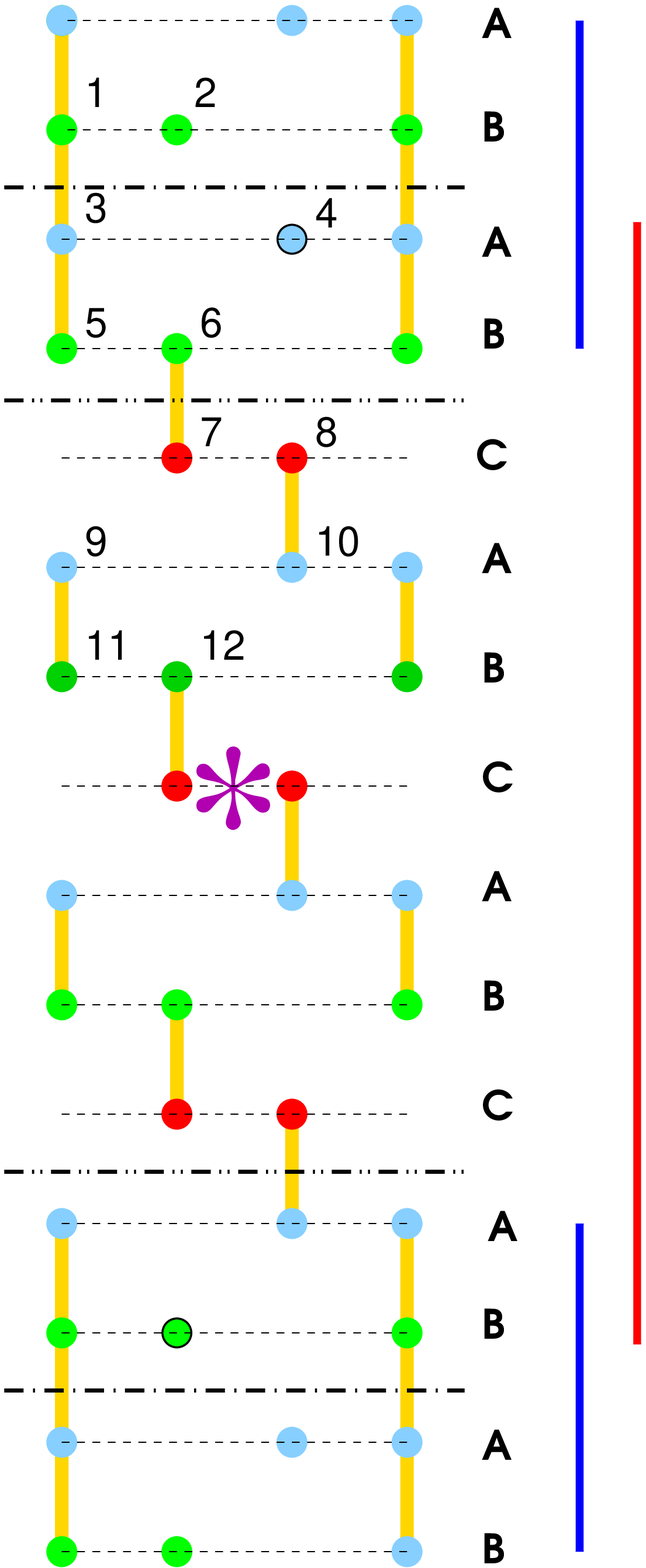}
\end{minipage}
\caption{ (Color online) Examples for the two types of interfaces 
in  super-cell geometry. 
Left panel: type $\alpha: (BA)_nC(ABC)_m(BA)_n$ and
right panel: type $\beta: (AB)_nC(ABC)_m(AB)_n$. 
The figure is for (n=2, m=2), but the numerical calculations are done with 
(n=3, m=4) and (n=6, m=8).
The numbering of the atoms is pinned to the interface, i.e., 
more layers for the numerical calculations are added outside the 
numbered region. 
The yellow perpendicular bars indicate overlapping $p_z$ orbitals \cite{note2}.
(Dangling $p_z$ orbitals ore not indicated.) 
The plot shows the atoms in the plane parallel to the paper plane of 
Fig.~\ref{fig:atoms-ABC}. 
(Note, that only the 3 left  vertical columns of atoms form one unit cell.) 
The vertical colored bars to the right of the figures indicate 
 the blocks of the two lattice types, 
the thick horizontal dash-dot and dash-double-dot lines mark geometrical 
borders of the bulk blocks (see text),  and 
the purple asterisks denote inversion centers. 
Atoms belonging to interface clusters are encircled.
The symmetry group is PB3M1 (164) in both cases.
(The figure is schematic and not true to scale.)
}
\label{fig:atoms-if}
\end{figure}

For our calculations of interfaces we used a super-cell geometry,
for which the unit cell is chosen such that
two interfaces
per unit cell and an inversion center in the middle are obtained (see
Fig.\ref{fig:atoms-if}).
The inversion symmetry  reduces the  numerical effort considerably
and prevents the formation of dipole moments.
Another possible option are  slabs
with one interface in the middle,
which are periodically repeated and separated by 
finite layers of vacuum.
In this case 
an inversion center (or a mirror plane) cannot be accommodated,
but the electronic structure of the surface could be studied
 together with the interface.
Because the surfaces have been investigated already in the former works
\cite{taut13} and \cite{taut14}, we preferred the super-cell model in this work, 
and use the slab model only for double-checking.
For establishing an inversion center, 
we added an additional C-layer on top of the (ABC) block. This trick 
makes the (ABC) block just one layer thicker, but does not 
disturb the stacking order.  
Please note, that the unit cells shown in the figure
are just for visualization and the real calculations are done with 
much thicker unit cells as described in the figure caption. 

As in our previous investigations of stacking faults, there are two types
of simple interfaces, which are denoted by $\alpha$ and $\beta$ and which
differ in their local bonding picture and consequently
 also in their electronic properties (see Fig.\ref{fig:atoms-if}).
Interfaces in graphite stacks can create {\em interface clusters}, which 
are not present in the adjacent bulk blocks. 
(A monomer is considered here as a special case of a cluster, and the term 
'monomer' and 'single atom' are considered as synonyms.)\\
In {\bf type} {\boldmath $\alpha$} the periodicity is broken at 
 a plane which can be located 
between atoms 5,6 and 7,8 (restricting ourselves to the upper half of the cell). 
Consequently, the 
 {\em trimer} (atoms 6, 8, and 10) 
and an additional {\em monomer} outside the (AB) block (atom 7) do not belong to 
the adjacent bulk structures and should be considered as interface clusters. \\
In  {\bf type} {\boldmath $\beta$}, the matter is not that easy. 
If we approach the interface: \\
(i) from the side of the (AB) block, the periodicity 
is broken at the plane between atoms 5,6 and 7,8, and atom 4 would 
be part of the (AB) block and therefore {\em no} interface cluster. \\
(ii) Alternatively, from the (ABC) side, the periodicity             
is broken at the plane   between atoms 3,4 and 1,2 and atom 4 does not belong 
to the (AB) block and consequently it   should be 
an interface  monomer.\\
Consequently, in this  case
 this issue cannot be decided on the basis of simple geometry. 
As we will see in the Section III.B, atom 4  hosts an interface state 
with the same characteristics as the 
interface state at atom 7 in type $\alpha$, 
and therefore atom 4 in type $\beta$ should be considered as an interface {\em monomer}.

\subsection{Interface $\alpha$}

\begin{figure}[h]
   \centering
 \includegraphics[width=0.9\textwidth]{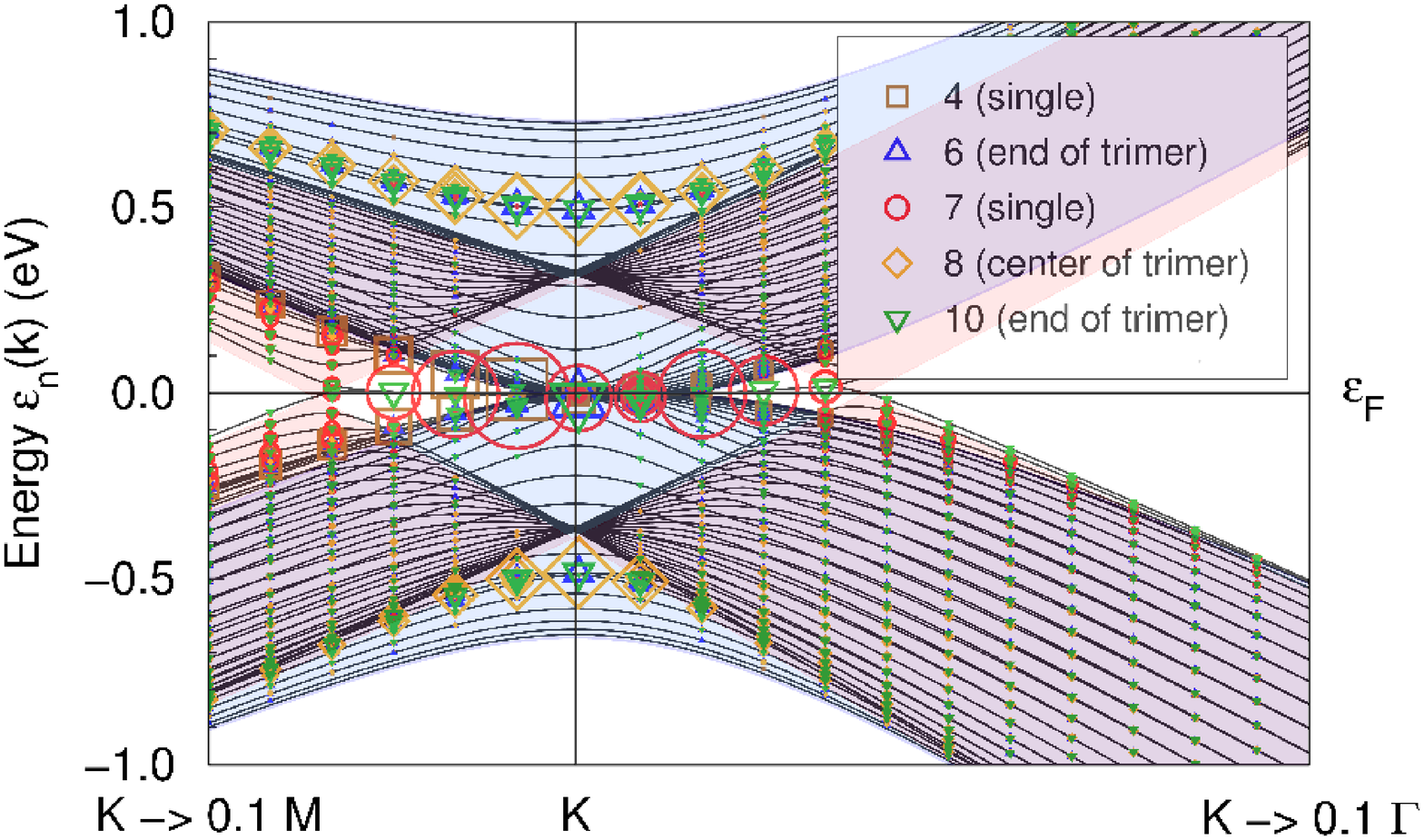}
 \includegraphics[width=0.9\textwidth]{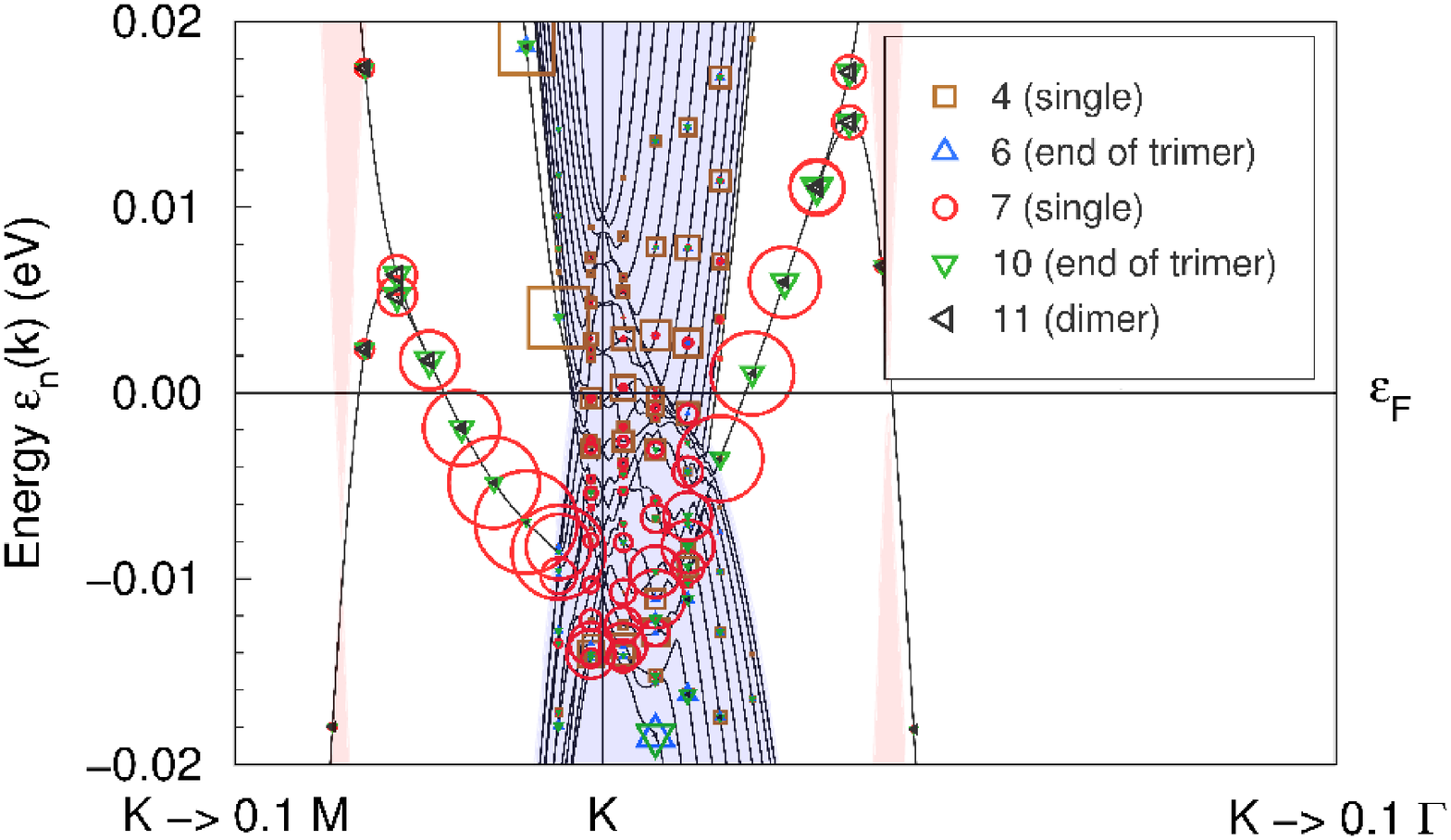}
 \caption{ (Color online)
 Band weights of   $2 p_z$-orbitals at some interface 
 atoms for  the $\alpha$-slab $(BA)_6C(ABC)_8(BA)_6$ 
are represented by the size of the corresponding symbols.
(A precise definition of the band weights can be found
 in the Introduction of Ref.~\cite{taut13}.)
The lower panel shows a zoom to energies close to the Fermi level.
For numbering of the lattice sites (in the figure caption)
see Fig.~\ref{fig:atoms-if}. 
The self-consistent band structure calculation 
has been done with a {\bf k} grid 
of 100x100x2 points.
}
   \label{fig:alpha-BW}
\end{figure}

 Fig.~\ref{fig:alpha-BW} shows the bands around the Fermi level
 $\varepsilon_F$ including band weights 
(for definition see Introduction of Ref.\onlinecite{taut13}) for certain atomic 
positions and along the symmetry line M-K-$\Gamma$ in the 
basal plane  around the K-point. 
Because the Brillouin zone of the super-cell 
is  about 65 times wider than high, the dispersion in ${k}_z$ direction is 
 practically not visible and therefore not shown. 
Note, that  in infinite 
super-cells all interface bands would be 
 doubly degenerate due to the occurance of 
two interfaces per unit cell. 
In our finite super-cells, however, these bands {\em can} split up, 
upon approaching the bulk 
state continuum, because then the decay length converges 
to infinity and  
 the  interface states  of the two interfaces overlap. 
This effect is clearly seen in the lower panel of Fig.~\ref{fig:alpha-BW} 
and Fig.~\ref{fig:beta-BW}, where the interface bands approaches the (red) bulk 
band continuum of (ABC)  graphite. 

As in the previous papers on  stacking faults in (AB) and (ABC) graphite, 
{\em some} of the  interface bands at the K point can be understood 
{\em qualitatively} using a simple model for the 
{\em  interface clusters}. 
In this model the atoms of the clusters 
are bound to each other only by one effective 
overlap integral $\gamma_1$=0.345 eV
perpendicular to the layers 
(see Appendix in Ref.~\onlinecite{taut14}).
The reason why the interface clusters are virtually  decoupled 
from the adjacent bulks is that at the K point the coupling 
between  atoms in the same layer cancels, and the interface clusters 
have no nearest neighbors in adjacent bulk layers 
(see e.g. Ref.~\onlinecite{guinea06}). 
In our case, this model provides  a {\em monomer}  state at E=0, 
and for the {\em trimer} one state at E=0 with eigenvector $(-1,0,1)$ and 
a pair of states at E=$\pm \sqrt{2} \;\gamma_1$ with eigenvectors 
$(1,\pm \sqrt{2},1)$. 
The first eigenstate is located  only at end-atoms of the trimers, whereas 
the pair of states has its main weight in the center. 

(i) The most prominent interface band 
related to an interface cluster is near the Fermi level  $\varepsilon_F=0$
 and  is mainly  localized at the {\bf monomers} (No.7, red circles) 
at the interface (see Fig.~\ref{fig:alpha-BW}, lower panel).  
In the window outside of both bulk continua, 
its wave function decays into both adjacent bulk blocks, 
whereas upon approaching the (ABC) continuum 
it hybidizes mainly with the end atom of the trimer No.10  
and dimer atom No.11 (not the much closer atom No.9 !) 
and correspondingly the band weight of  atom No.7 decreases. 

(ii) A pair of  interface bands  is localized  at the {\bf trimer} (atoms
 No. 8, 6 and 10, blue and green triangles and golden diamonds)
  with energies around $\pm$ 0.5 eV. 
Closer examination of the weights of further atoms than those shown in 
Fig.~\ref{fig:alpha-BW} reveals that these trimer bands are localized 
in both directions perpendicular to the interface,
 despite the fact that they are located within the 
(AB) bulk continuum. Hybridization of interface and bulk states, 
which share the 
same region in the E - $k_{||}$ space {\em may} take place, but  
can be suppressed or reduced in certain symmetry or binding configurations.

\begin{figure}[h]
   \centering
 \includegraphics[width=0.9\textwidth]{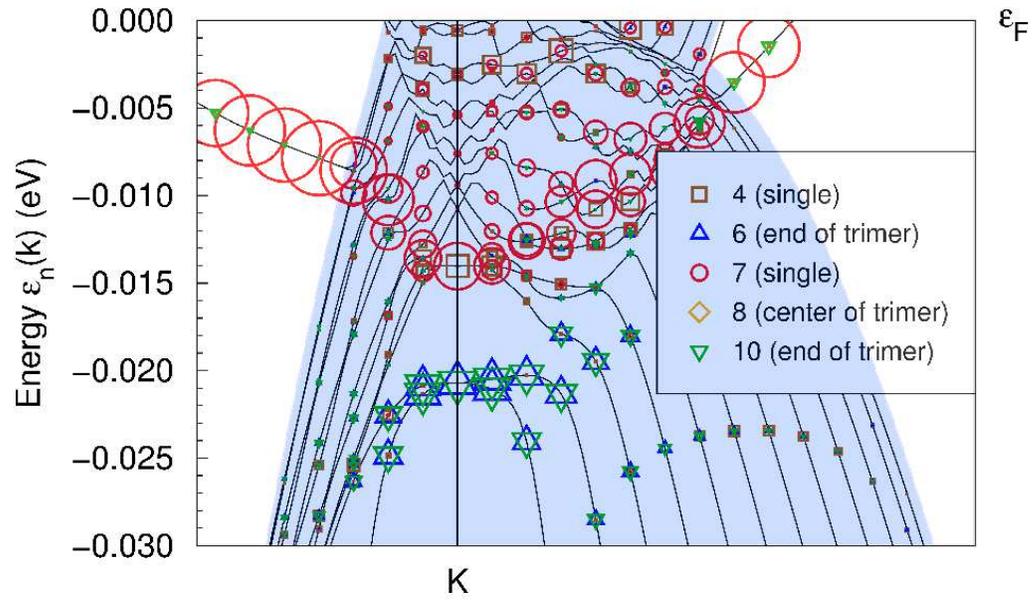}
 \caption{ (Color online)
Zoom of Fig.\ref{fig:alpha-BW} to the energy - $\bf k$ window
of the  trimer state.
}
   \label{fig:alpha-trimer-BW}
\end{figure}

(iii) The third {\bf trimer} state  with energy at the K-point around -0.02 eV
is visible in the  special zoom shown in Fig.\ref{fig:alpha-trimer-BW}.
 Its  wave function is located at the ends  of the trimer 
(atoms No. 6 and 10, blue and green triangles).
Like the pair of trimer bands discussed in (ii), 
it is localized in both directions 
perpendicular to the interface.

There is a  different type of interface band, however, which is localized 
at the {\bf monomer} No.4 (brown squares), 
which is {\em not} an interface cluster, 
 but it is  the 
monomer within the (AB) bulk structure closest to the interface. 
The dispersion of this band differs completely from the 
{\em interface-cluster-induced} bands. Therefore we call it 
{\em surface-band-like}.
This interface band   runs  close and almost parallel to the boundary
of the (AB) bulk continuum and it resembles the interface bands
near a {\em stacking fault} in (AB) graphite 
reported in Ref.~\onlinecite{taut13}.
Note, that this band is not restricted to positive energies as it seems 
in the lower panel of Fig.~\ref{fig:alpha-BW}. 
Only in the energy window between -0.02 and 0 it is suppressed by the 
strong interface band located at atom No.7 (red circles).
 Closer examination of the upper panel reveals that it reappears  
in the energy range below -0.02.
However, this band is strictly localized 
 only on the line K $\to$ 0.1 M (left of the (AB) continuum), 
but  on K $\to>$ 0.1 $\Gamma$ 
 it lies within the continuum and turns into an interface  resonance. 

\begin{figure}[h]
   \centering
   \includegraphics[width=0.81\textwidth]{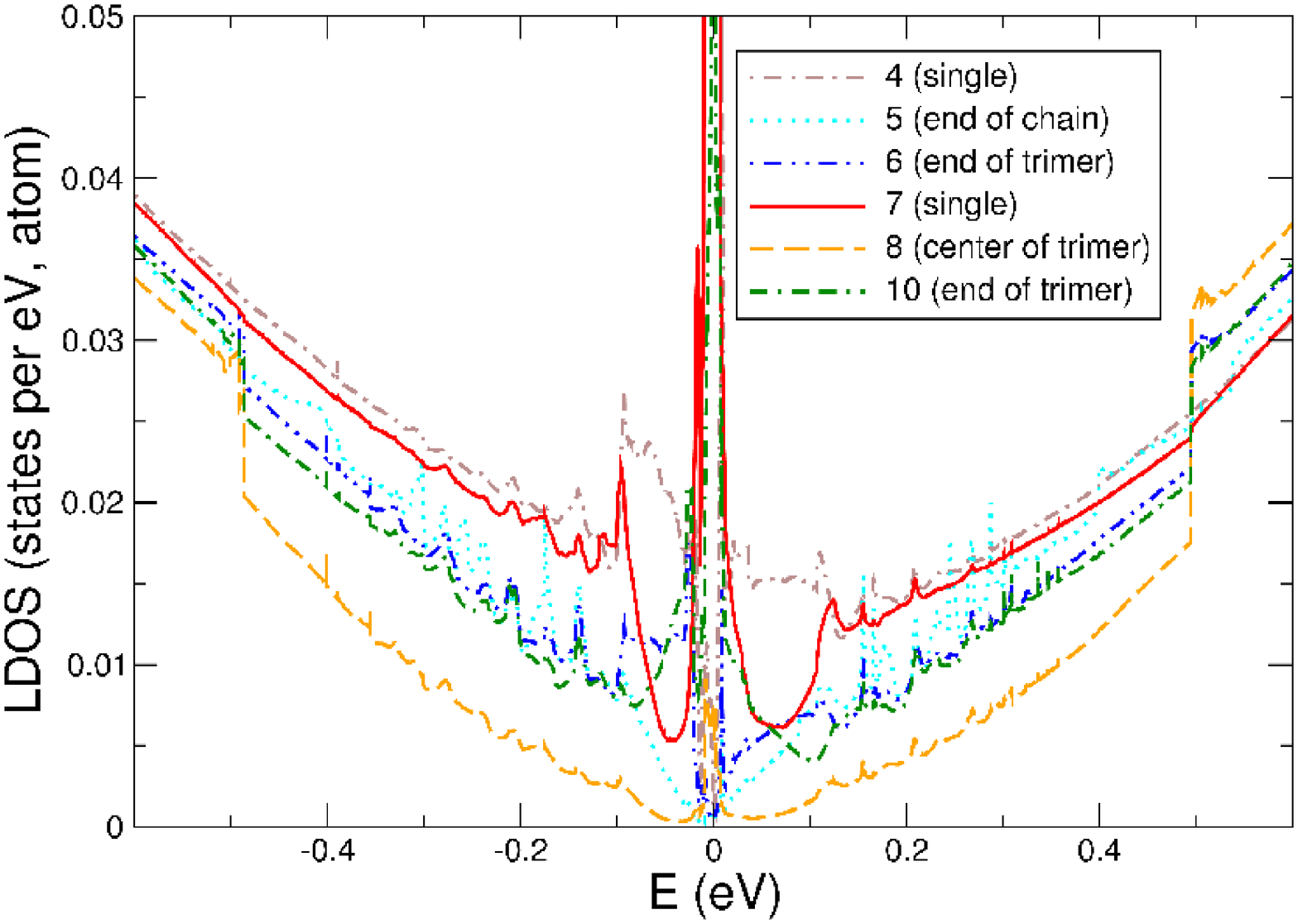}
   \\[4mm]
   \includegraphics[width=0.8\textwidth]{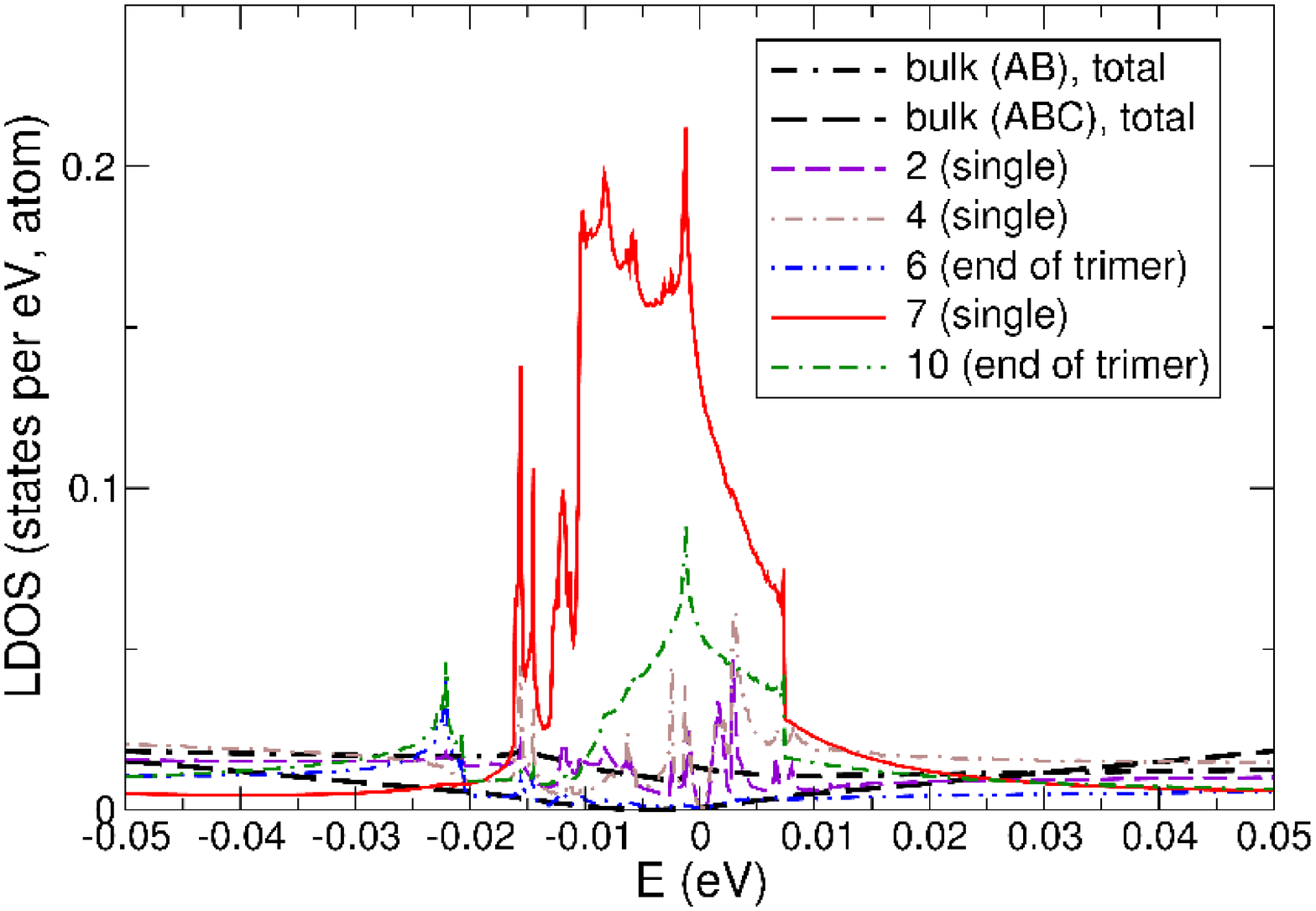}
   \caption{ (Color online)  
Local density of states (LDOS) for the most important sites close 
to the interface 
in the $\alpha$-slab $(BA)_3C(ABC)_4(BA)_3$.  
The {\bf k} grid for the {\bf k}-space integration 
in the upper panel consists of 250x250x5 points in the whole BZ. 
The zoomed version in the lower panel was 
calculated using  a special grid consisting
of 100x100x3 points closely  localized around the symmetry line K-H. 
(For numerical reasons, the LDOS had to be calculated with a thinner slab as the 
band structure.)
}
   \label{fig:alpha-LDOS}
\end{figure}

In Fig.~\ref{fig:alpha-LDOS} the LDOS is presented for a selection of atoms 
on two different energy scales.  The most prominent peak 
at the Fermi energy comes from the interface band at the monomer 
(single atom No.7).  
 The next larger contribution to the LDOS at the Fermi energy 
 is located at  one end of the trimer (atom No.10) and stems
 from the low-energy trimer band.
 The high-energy trimer bands (at atoms 6, 8, and 10) around $\pm$ 0.5 eV, 
which have  a 2D dispersion of type $k_x^2 + k_y^2$, 
 produce only  the steps in the LDOS seen in the upper panel.

\subsection{Interface $\beta$}

\begin{figure}[h]
   \centering
 \includegraphics[width=0.9\textwidth]{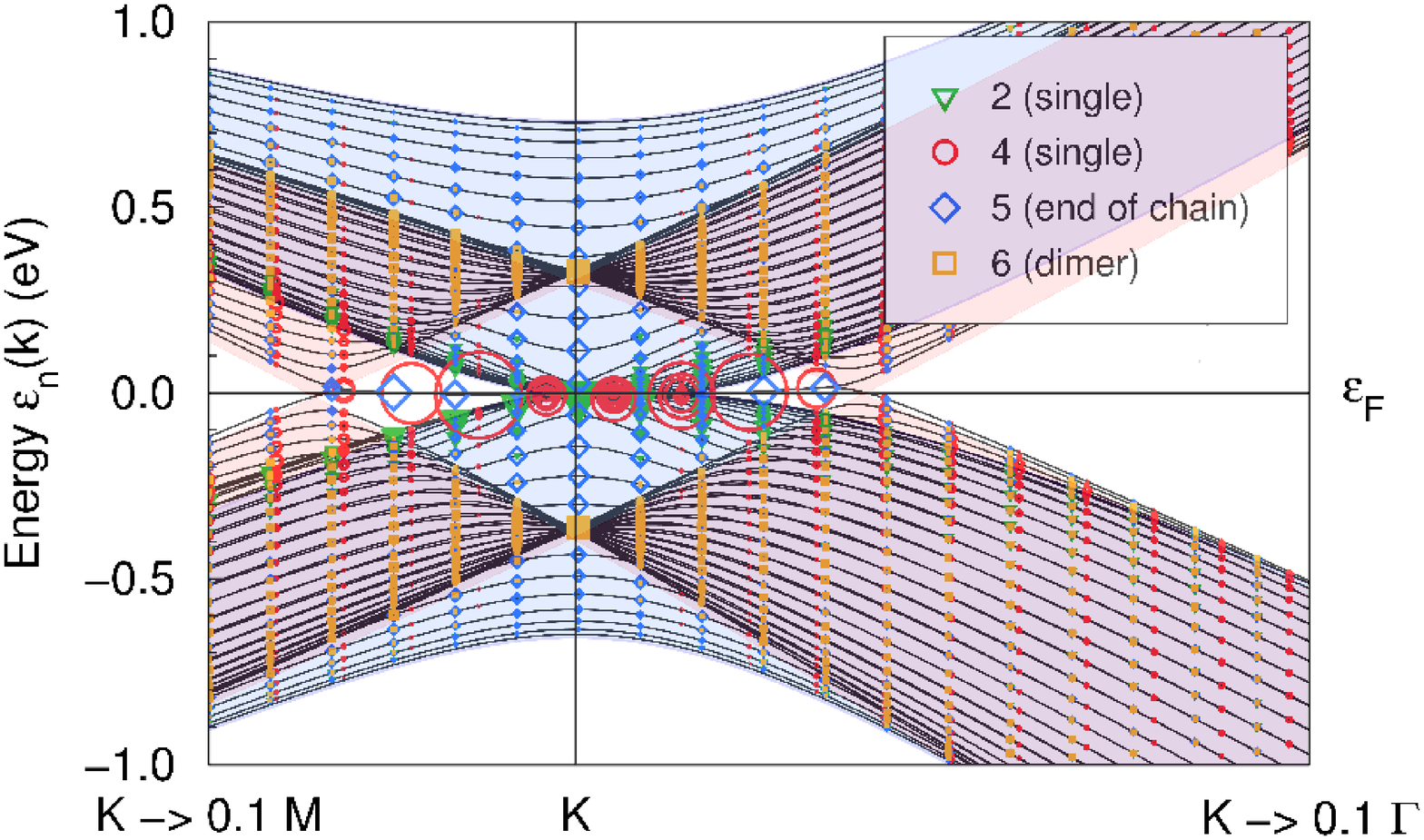}
 \includegraphics[width=0.9\textwidth]{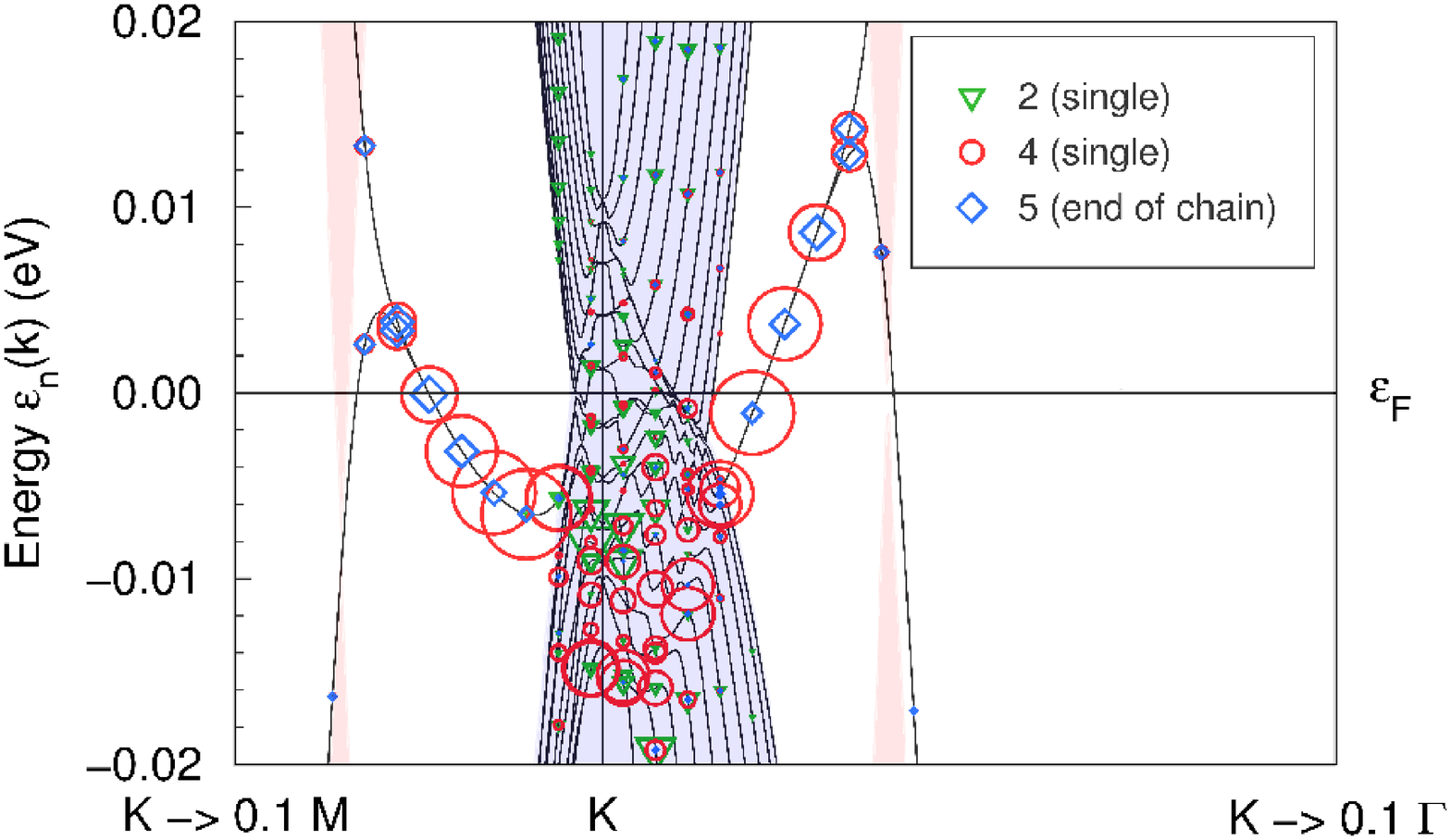}
 \vspace{-1cm}
 \caption{ (Color online) Band weights of   $2 p_z$-orbitals at some 
 interface atoms  for  the $\beta$-slab $(AB)_6C(ABC)_8(AB)_6$.
All  other details and conventions are analogous to 
 Fig.~\ref{fig:alpha-BW}.
}
   \label{fig:beta-BW}
\end{figure}

\begin{figure}[h]
   \centering
   \includegraphics[width=0.81\textwidth]{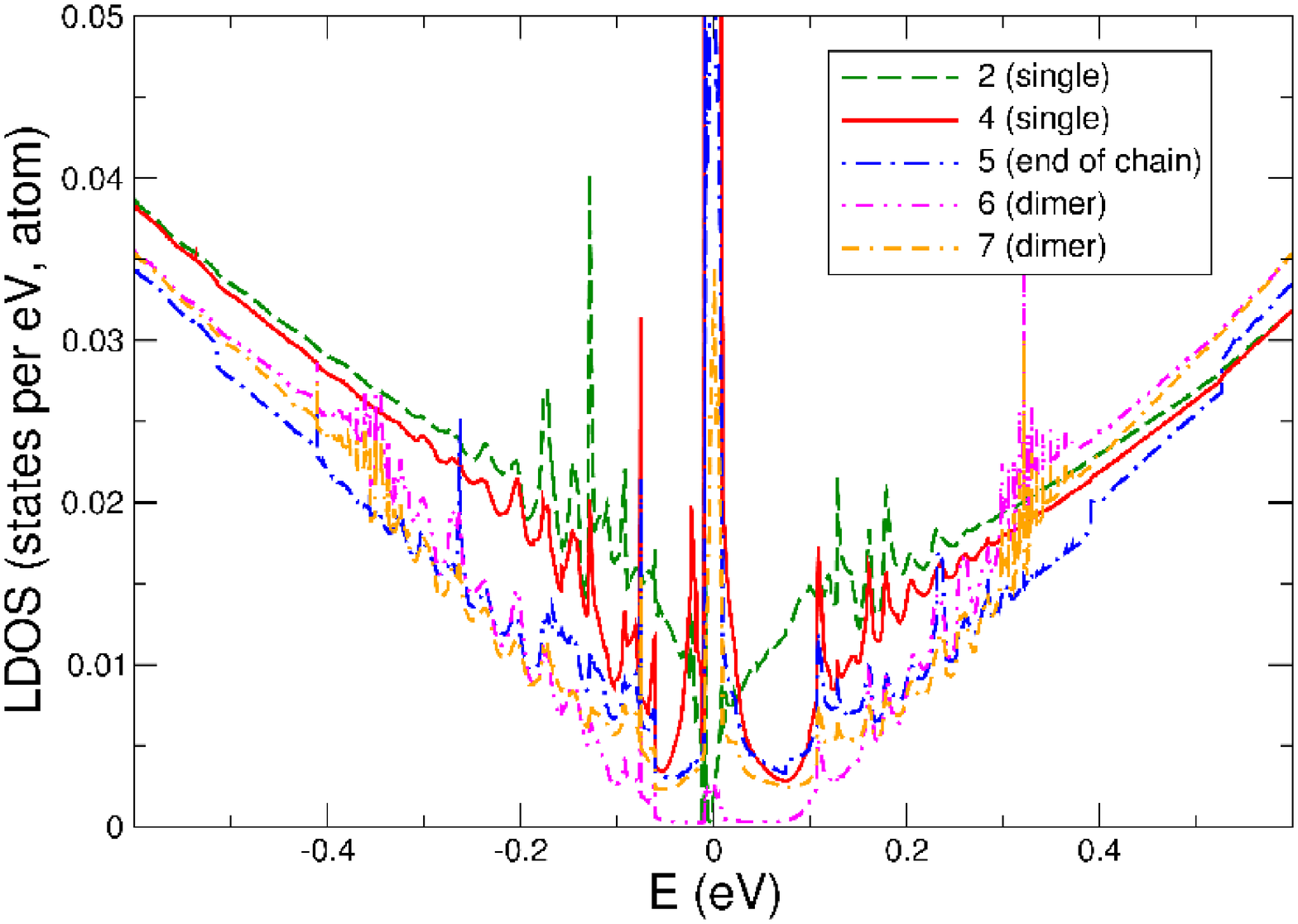}
   \\[4mm]
   \includegraphics[width=0.8\textwidth]{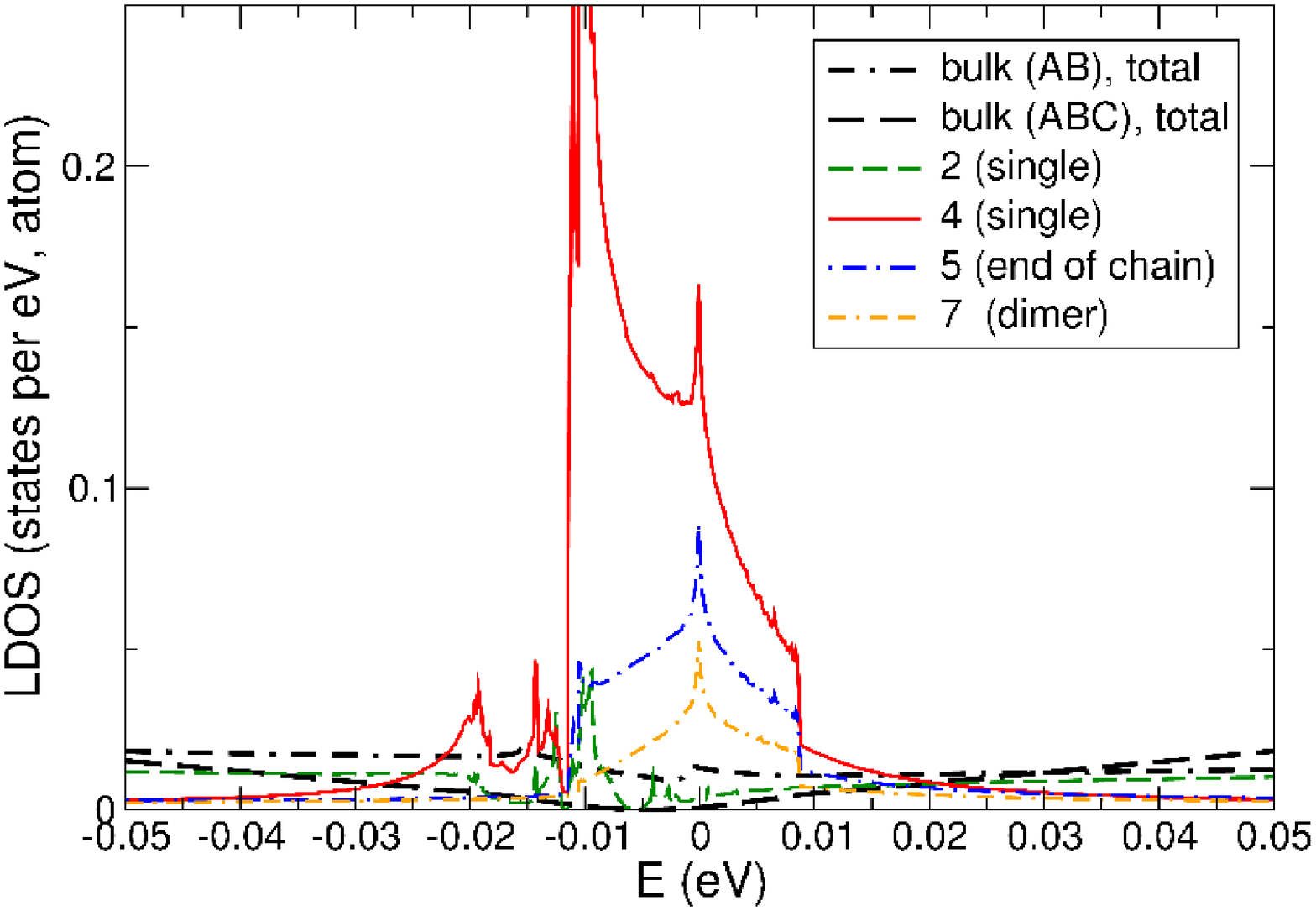}
   \caption{ (Color online) 
Local density of states (LDOS) for the most important sites close to the interface
 for  the $\beta$-slab $(AB)_3C(ABC)_4(AB)_3$.
The {\bf k}-grid is the same as in Fig.~\ref{fig:alpha-LDOS}.
}
   \label{fig:beta-LDOS}
\end{figure}

The bands in slab $\beta$ shown in Fig.~\ref{fig:beta-BW}
 are less rich in interface bands  because 
there is only one {\bf monomer} which can be considered as an interface cluster. 
The corresponding interface band is located at atom 4 
and its dispersion is very similar to the band located at atom 7 in type $\alpha$. 
 Correspondingly,
 the LDOS at the Fermi energy shown in Fig.~\ref{fig:beta-LDOS} is dominated 
by the peak from this band. Smaller peaks are from atoms 5 and 7.  
The contribution of atom 6 (not shown)  is 
much smaller than the contribution of the atoms included in the figure, 
even though it is located close to the interface as well and belongs to the same dimer.
 The latter 
effect is similar to that in type $\alpha$, where the contribution of 
atom 9 is much smaller than atom 11. 
In type $\beta$ we do not have a {\em surface-band-like}
 interface band which is the equivalent 
of the band at atom 4 in  $\alpha$, i.e., a band which is located at 
the monomer within (but close to the border of) the (AB) bulk block.
 This shows that the {\em  surface-band-like}  interface bands 
are very sensitive to details of the structure, as well known from non-topological 
surface bands.

\section{Interface energy}
The interface energy per interface atom is defined
 in analogy to formula (7) in Ref.~\cite{taut13} 
\begin{equation}
 e_{if} = \frac{1}{4} \bigg (
 E_{sc} - N_{at} \frac{(e_{hex} + e_{rhomb})}{2} \bigg )
\label{defe_if}
\end{equation}
where $E_{sc}$ is the total energy of the super-cell with $N_{at} $ atoms, 
and $e_{hex}$ and $e_{rhomb}$ are the total energies per atom for
 hexagonal and rhombohedral bulk graphite. 
The factor $1/4$ considers the fact that we have two interfaces  per unit
cell and two atoms in each interface layer.
Unlike in the case of stacking faults, however, we cannot subtract in  (\ref{defe_if})
the ground state energy of the supercell without interface, 
because this definition would lead to a quantity which 
is roughly proportional to the distance of the interfaces.
Instead we subtract the total energy of a fictitious system, 
in which the two bulks extend upto the mathematical interface plane. 
This provides a quantity which is (in the limit of thick supercells)
  independent of  thickness. 
Note, however, that $e_{if}$ is not the formation energy of the interface, 
but it can be used to find the most probable interface structure. 
As seen in Table~\ref{tab1}, structure $\beta$  
is energetically much more favourable than $\alpha$. 
Comparison with the data      on hexagonal graphitic stacks
 given in Ref.~\cite{taut13} shows that 
the {\em interface} energy between hexagonal and rhombohedral graphite presented here 
is one order of magnitude larger than {\em  stacking fault} energies, but one to two 
orders smaller than {\em surface} energies. On the other hand, the energy differences 
 given in Table~\ref{tab1} are of the same order as the difference in the 
total energies (per atom) of {\em bulk} hexagonal and rhombohedral graphite given 
in Ref.~\cite{taut13} as well.

\begin{table}[h]
\begin{tabular}{|c|c|r|}
\hline
system & formula & $e_{if}$ \\
\hline
$\alpha$ & $(BA)_3 C (ABC)_4 (BA)_3$  &  257.1   \\
\hline
$\beta$  & $(AB)_3 C (ABC)_4 (AB)_3$    &   64.2 \\
\hline
\end{tabular}
\caption{
Interface energy per interface atom $e_{if}$, as
defined in the text, (in $\mu$eV)  in the LDA. The $\bf k$-grid for the 
supercell calculations is 100x100x3 and 
for the overlap matrix elements the increased precision \cite{note1} 
is applied throughout.
}
\label{tab1}
\end{table}


\section{Summary}
A central role for the interpretation of interface bands is played by 
{\em interface clusters}. They are  caused by the interface and
 do not exist in the adjacent bulk materials. In {\em  this series of papers} we met 
monomers, dimers,
linear trimers, and linear tetamers. Monomers and trimers
 host interface bands around the Fermi energy, and dimers,
trimers and tetramers {\em also} 
at finite energies in the range between $\pm$0.2 and $\pm$0.5 eV.
(Please note, that monomers and dimers can also belong to the bulk blocks
of (AB) and (ABC) graphite, respectively. In this case they are not called 
'interface clusters'.)

In {\em this paper} we  investigated the two simplest
 types of interfaces between hexagonal (AB) 
and rhombohedral (ABC) graphite: type $\alpha$ hosts  two different 
 interface clusters, namely trimers  and monomers, whereas 
 type $\beta$ has only interface monomers.
From  total energy calculations it follows that 
type $\beta$ has the lower interface energy and it is therefore 
more probable to occur. 
Both interfaces produce  interface bands near the K point in the Brillouin zone.  
These interface bands can be categorized into two two groups: \\
(i) Most of the interface bands can be traced back to interface clusters. 
In type $\alpha$, they are localized at atom 7 (monomer) and atoms 6, 8, and 10 (trimer),
and in type  $\beta$ at atom 4. \\
(ii) However,  there is also the possibility 
for surface-band-like interface bands, 
which are localized at the border, but within the adjacent bulk materials. 
In type $\alpha$, the band localized at atom 4 (which belongs to the (AB) block) 
 is such a case. 
As in the case of non-topological surface states, the existence 
 of the second type of interface bands depends
 on the details of the structure and potential near the surface/interface. 
Interface $\beta$ hosts none of this type. 

Interface bands near the Fermi energy produce large peaks in the LDOS, 
which in turn should lead to large contributions to the local conductivity 
along the interfaces. The trimer bands around $\pm 0.5$ eV cause only steps 
in the LDOS. 

\begin{acknowledgments}
We are indebted to M. Richter 
 for support and helpful discussions.
\end{acknowledgments}


\bibliographystyle{apsrev}  
\bibliography{ref.bib}

\end{document}